\documentclass[journal=jacsat,footinbib,manuscript=article]{achemso}
\usepackage{amsmath}
\usepackage{amssymb}
\usepackage{dcolumn}
\usepackage{graphicx}
\usepackage{longtable}
\usepackage{bm}
\usepackage{color}
\usepackage{epsfig}
\usepackage{hyperref}

\usepackage{chemformula} 
\usepackage[T1]{fontenc} 

\author{Gang Bahadur Acharya$^{1,2\dag}$}
\author{Mohan Bikram Neupane$^{1\dag}$}

\author{Rojila Ghimire$^{1}$}

\author{Madhav Prasad Ghimire$^{1}$}

\email{madhav.ghimire@cdp.tu.edu.np}
\affiliation[Tribhuvan University]
{$^1$Central Department of Physics, Tribhuvan University, Kirtipur, 44613, Kathmandu, Nepal\\
	$^2$Leibniz IFW-Dresden, Helmholtzstr. 20, D-01069 Dresden, Germany}

\title
  {Weyl Metal Phase in Delafossite Oxide PtNiO$_2$ }

\begin{document}
\begin{abstract}
 On the basis of density functional theory calculations we  predict Weyl points in rhombohedral structure of PtNiO$_2$ having symmorphic symmetry. From the formation energy and phonon calculations, PtNiO$_2$ is found to be structurally stable.
 The magnetic ground state is ferromagnetic with an effective magnetic moment of 1.01 $\mu_B$ per unit cell.
 The electronic structure shows major contributions from Pt-$5d$, Ni-$3d$ and O-$2p$ orbitals with band crossing close to the Fermi level. The orbital contribution around 8 eV above the Fermi level are from the Pt-$s$, $p$ orbitals forming a hidden kagome-like electronic structure confirmed by surface Fermi surface spectral function.
 We found 20 pairs of confirmed Weyl nodes along the magnetic easy axis [100]. These results are expected to provide a useful and exciting platform for exploring and understanding the magnetic Weyl physics in delafossites.

\end{abstract}
{\footnotetext[1]{\dag~These authors contributed equally to this work.}}

\section{Introduction}
Materials with unusual quantum phenomena, such as unique transport features and topological surface states, topological semimetals, etc. have generated  lots of interest in recent years \cite{RevModPhys.90.015001}. In electronic band structures of topological semimetals, the interaction of symmetry and band topology is important. As a result, topologically protected zero-dimensional point-like or one-dimensional line-like Fermi surfaces in momentum space are formed \cite{PhysRevLett.122.057205}. Topological semimetals emphasize various forms of low-energy excitations near the protected band-crossing points, such as Dirac fermions \cite{PhysRevB.85.195320,doi:10.1126/science.1245085}, Weyl fermions \cite{PhysRevB.83.205101,PhysRevLett.107.127205,PhysRevLett.107.186806,PhysRevLett.115.265304,soluyanov2015type}, nodal line fermions \cite{PhysRevB.84.235126}, and triple fermions \cite{PhysRevX.6.031003,Winkler_2019}, under various symmetries in the crystals. Weyl semimetals (WSMs) significate at extension since  both the conduction and valence bands cross each other near the Fermi level ($E_{ F}$), and the crossing points, known as Weyl points (WPs), appear as monopoles of Berry curvature in the momentum space. WPs have a definite chirality  +1 or -1. The Berry curvature is a magnetic field in the momentum space that causes anomalous velocity of electron motion in real space.
As a result, Weyl monopoles play an important role in electrical conduction, such as in the anomalous Hall effect. WPs  location and their energies are significantly influenced by magnetic order as well as spin structures~\cite{yan2021weyl}. Using spectroscopic and electrical transport approaches, many WSMs have already been identified theoretically and  confirmed experimentally~\cite{RevModPhys.90.015001}. The majority of these compounds are nonmagnetic, therefore achieving  WPs require the presence of a non-centrosymmetric crystal structure. Nonmagnetic WSMs have lots of charge carrier mobility and a significant  magnetoresistance~\cite{shekhar2015extremely,kumar2017extremely}. If the compound is magnetic, the WPs can appear even in centrosymmetric structures~\cite{doi:10.1021/acs.chemmater.9b05009}. Magnetic WSMs have recently attracted attention as a new platform for exploring the interaction of chirality, magnetism and topological order. These studies  potentially lead to novel quantum states, spin-polarized chiral transport~\cite{hosur2013recent}, and unusual optical features~\cite{PhysRevB.94.245121,PhysRevB.95.085127,PhysRevLett.122.197401}. Compared with the non-magnetic WSMs, magnetic WSMs gives us a powerful tool for manipulating band structure and associated electromagnetic performance. Magnetic arrangement breaks the time reversal symmetry in magnetic WSMs, requiring additional symmetry to protect the topological property. Pyrochlore iridate Re$_2$Ir$_2$O$_7$ (Re=rare earth) \cite{PhysRevB.83.205101} and  HgCr$_2$Se$_4$ \cite{PhysRevLett.107.127205}, for example, are the first magnetic topological WSMs proposed, and they have several pairs of time-reversal-symmetry-breaking Weyl nodes. A half-metallic ferromagnet  Co$_3$Sn$_2$S$_2$ \cite{liu2018giant,wang2018large,doi:10.1126/science.aav2873,PhysRevResearch.1.032044}, is also predicted as  magnetic WSM which contains only six WPs above $E_{ F}$. From density functional theory (DFT) calculations and  using angle-resolved photoemission spectroscopy experimentally, WSMs were found. In magnetic WSMs, there are possibilities for the creation, annihilation and shifting of WPs due to the effect of magnetization rotation \cite{PhysRevResearch.1.032044}. Also, Co$_2$MnGa \cite{sakai2018giant,guinnpg}, Co$_2$MnAl \cite{kubler2016weyl}, Mn$_3$Sn/Mn$_3$Ge \cite{kuroda2017evidence,Yang_2017}, GdPtBi \cite{hirschberger2016chiral} and YbMnBi$_2$ \cite{borisenko2019time} are demonstrated as  magnetic WSMs. \\
After 1997, the delafossite structural series of oxides ABO$_2$, where (A=Pt, Pd, Ag and Cu and B=Cr, Co, Fe and Ni) \cite{doi:10.1021/ic50098a011,doi:10.1126/sciadv.1500692,doi:10.1021/ic50098a013,cerqueira2018structural} have been studied due to the transparent conductive property found in Cu based delafossites \cite{cerqueira2018structural}. Likewise, Pt and Pd based delafossites having \textit{d$^9$} configuration possess high conductivity, comparable to that of Cu and Ag \cite{daou2017unconventional}. In search of PtCoO$_2$'s high conductivity,
it  was reported that   6\textit{s} and 6\textit{p}$_x$/\textit{p}$_y$ and \textit{d}$_{3z^2-r^2} $+ \textit{d}$_{xy}$/\textit{d}$_{x^2-y^2} $ orbitals of Pt formed hidden kagome-like features \cite{PhysRevMaterials.3.045002}. PdCoO$_2$'s high mobility was also compared to that of NbAs (a Weyl semimetal),\cite{zhang2019ultrahigh} implying that metallic  delafossites can be an alternative to today's semi-conducting devices. But, hitherto topological features of delafossite structures are not investigated. Thus, our work paves a pathway for the realization of Weyl features in the magnetic delafossites which has never been explored.  Our particular interest is in searching for the topological Weyl properties in delafossite structural oxide PtNiO$_2$. \\
By means of DFT calculations, we show that ferromagnetic delafossite oxide PtNiO$_2$ is a magnetic WSM. The electronic, magnetic and Weyl characteristics of delafossite oxide PtNiO$_2$ are investigated. The gap between the conduction and valence band opens due to the effect of spin orbit coupling (SOC), resulting in topological bulk Weyl properties as well as its surface properties. Along the [100] magnetic easy axis, we found total of 40 Weyl points. The proximity of all the WPs  to $E_{ F}$  suggests the possibility of  transport experiments. Our findings provide an excellent material for additional experimental synthesis and investigations of time-reversal breaking ferromagnetic WSMs.
 \section{Methods}
We performed DFT calculations to investigate the electronic properties of PtNiO$_2$ using the full-potential local orbital code (FPLO) \cite{PhysRevB.59.1743}, version 18.00-52, with a localized atomic basis and full potential treatment. The exchange-correlation energy functional used is the generalized gradient approximation in the parameterization of Perdew, Burke, and Ernzerhof's (PBE - 96) \cite{PhysRevLett.77.3865}. Using the PYFPLO module of the FPLO tool \cite{PhysRevB.59.1743}, we extract the Wannier tight-binding Hamiltonian by projecting Bloch states over atomic orbital-like Wannier functions for further analysis of the electronic structure. The localized Wannier basis states include  Pt [$5d, 6s$], Ni [$3d, 4s$], and O [$2s, 2p$] orbitals. The Wannier model is converged in the Brillouin zone (BZ) with a 12 $\times$ 12 $\times$ 12 grid sample. These Wannier Hamiltonians are then used to investigate the Weyl properties along various magnetization directions, as well as associated topological aspects including  surface attributes.  The energy convergence criterion was chosen at $10^{-6}$ Hartree. Dynamical stability has been studied using density functional perturbation theory (DFPT) implemented in quantum espresso code \cite{giannozzi2009quantum}. Broyden-Fletcher-Goldfarb-Shanno (BFGS) method has been used to perform ground-state optimization. The pseudopotential used is non-conserving type. The plane-wave basis set cut off energy chosen is 95 Ry. For Brillouin zone integration 12 $\times$ 12 $\times$ 2 \textit{k} mesh was taken. The Kohn-Sham equation is solved using the iterative Davidson-type diagonalization approach with an energy convergence threshold of 10$^{-10}$ Ry. We have used 4 $\times$ 4 $\times$ 1 \textit{q} mesh which calculates the phonon at any wave vector. 
The surface states in the projected two-dimensional (2D) BZ were obtained from the surface Greens function of the semi-infinite system. \cite{PhysRevB.96.195309}
\section{Results and discussion}
\subsection{Crystal structures}
\begin{figure}[h]
	\centering

 \psfig{figure=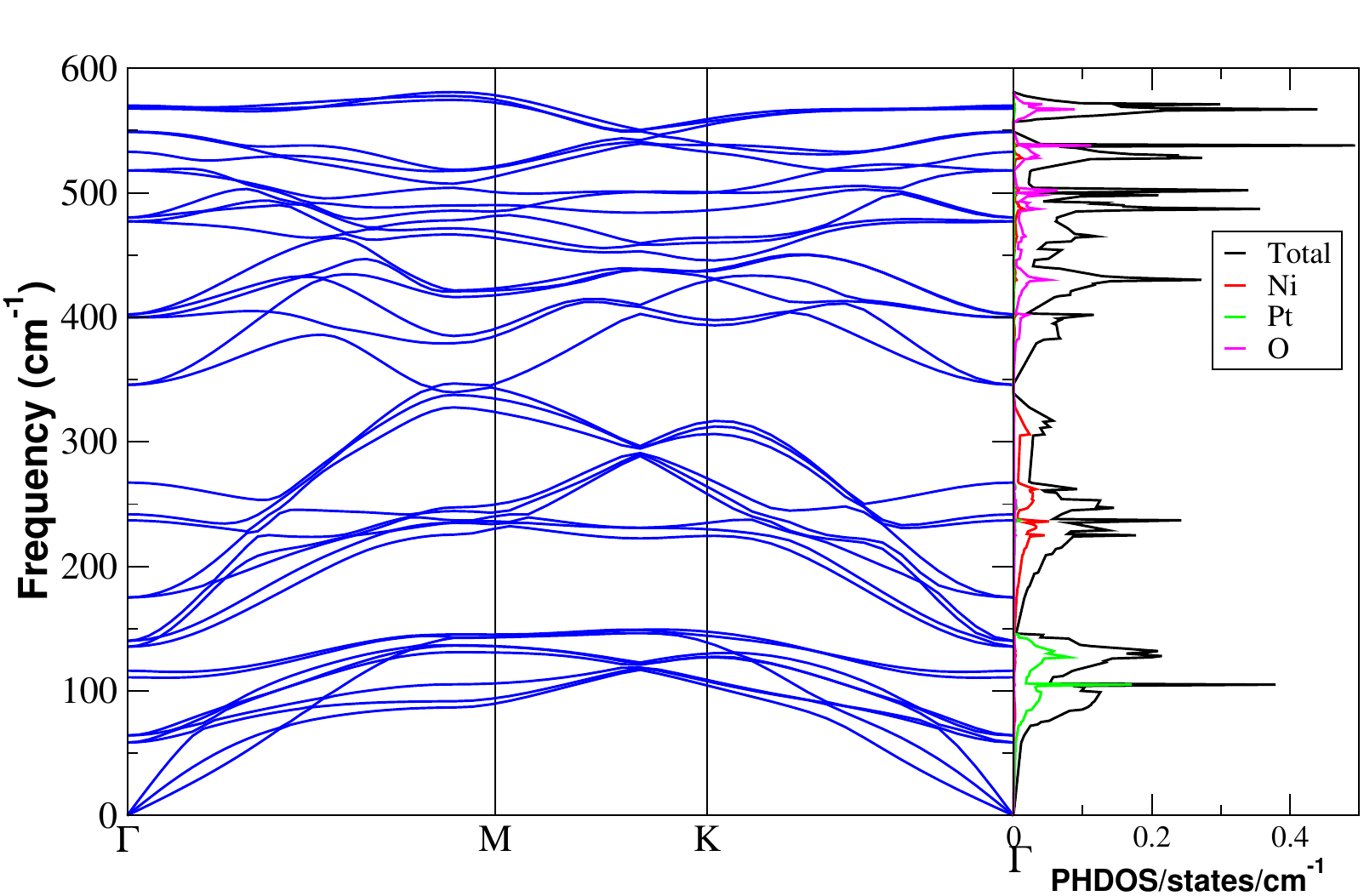,width=\columnwidth,height=4in}
	\caption{Calculated phonon spectrum of PtNiO$_2$}\label{1}
\end{figure}

PtNiO$_2$ belongs to a trigonal system of hexagonal crystal family with space group \emph {R$\bar{3}$m} (no. 166) \cite{jain2013commentary}. The structure is quasi two dimensional with lattice parameters \emph a = \emph b = 2.92 \AA, \emph c = 18.24 \AA  \hspace{ 0.2cm}and angles $\alpha = \beta$  = 90$^\circ$ and $\gamma$ =120$^\circ$. Platinum cation is linearly bonded with two oxygen anions. Nickel forms octahedron with oxygen. This delafossite comprises of Pt layer stacked between NiO$_6$ octahedra. The point group is $\bar{3}$ or D$_{3d}$. It is a symmorphic system with inversion center.
\begin{figure}[h]
	\centering
	
	\psfig{figure=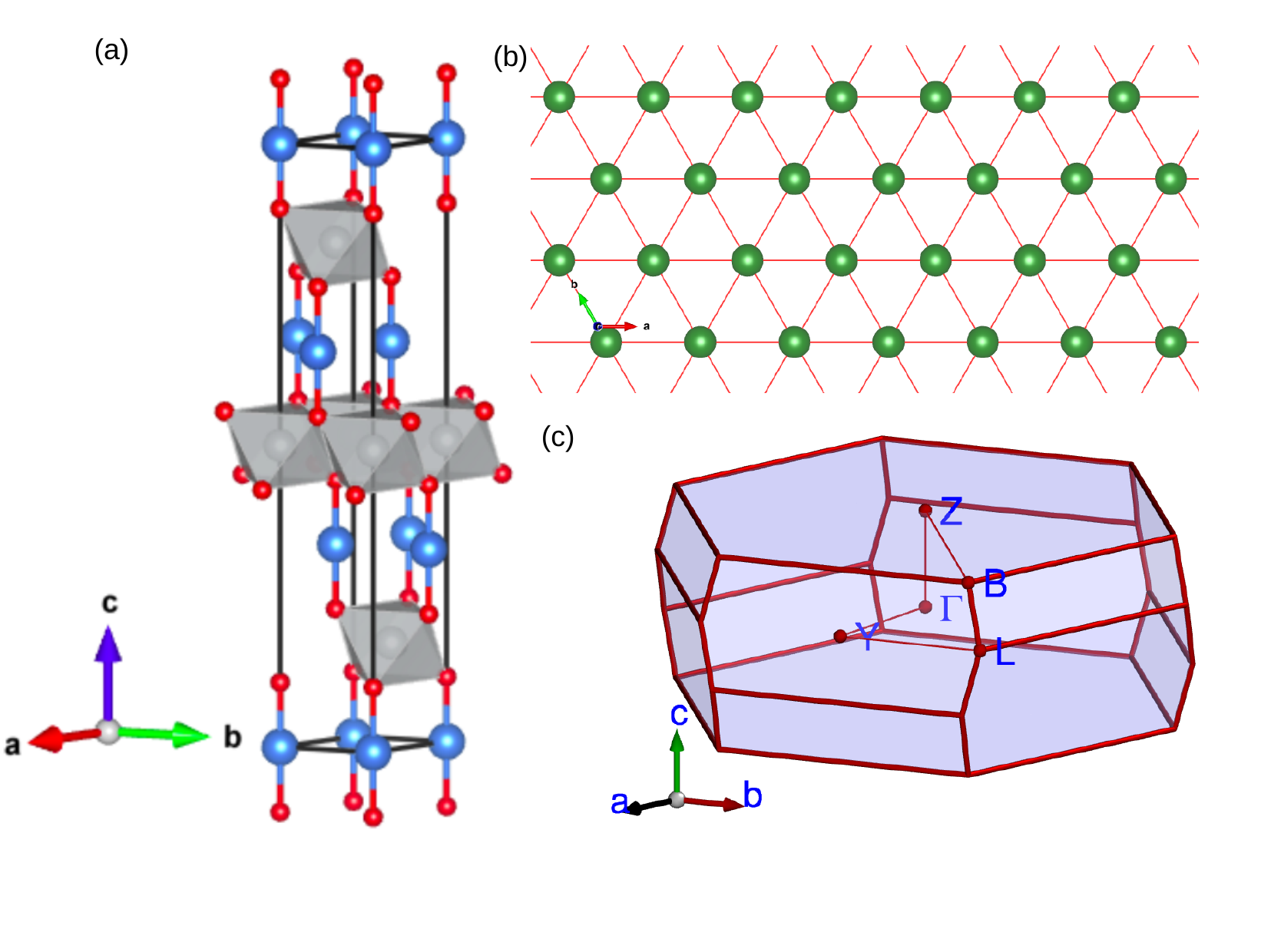,width=\columnwidth,height=4in}
	\vspace{-0.5cm}
	\caption{(a) Crystal structure of PtNiO$_2$ (blue, red and silver  balls represent Pt, O, and Ni atoms respectively,  octahedral arrangement between Ni and O atoms); (b) The triangular lattice of Ni atoms in a-b plane and (c) Bulk BZ. The red dots and lines indicate the high-symmetry points and paths respectively in the BZ.}
	\label{crystal}
\end{figure}
 In order to examine the stability of PtNiO$_2$, the formation energy was  calculated  using given equation:
\begin{equation}
E_{For} = E_{PtNiO_2}^{Tot} - [E_{Ni}^{bulk} + E_{Pt}^{bulk} + 2E_O^{bulk}]
\end{equation}
where $E_{PtNiO_2}^{Tot}$ indicates the total energy of the crystal. $ E_{Ni}^{bulk}$, $E_{Pt}^{bulk}$, and $E_O^{bulk}$ are the total energies of nickel, platinum and oxygen atoms respectively obtained from the bulk energy calculations.
The formation energy per atom is -0.58 eV. The negative value of formation energy shows the thermodynamical stability of  PtNiO$_2$.\\
The calculated phonon dispersion curve is shown in Fig. \ref{1}. The unit cell contains 12 atoms, resulting in 36 phonon dispersion modes with three acoustic and  remaining optical branches. 
The partial phonon density of states are presented in Fig. \ref{1} which depicts the major contribution of O and minor contribution of Ni in higher optical modes whereas the middle optical modes are mainly due to O. The frequencies of the acoustic phonon branches are in an interval of 0  up to  110 cm$^{-1}$, with an overlap with optical phonon modes of low frequencies. The higher optical modes are less dispersive than optical modes at lower and middle regions which represents strong intra-molecular interactions. At high symmetry point M, O atom vibrate with frequency 581 cm$^{-1}$, which is maximum frequency of vibration.   The real phonon spectrum shown Fig. \ref{1}  ensures the dynamical stability of the system and provides an insight for experimental realization.  
\subsection{Electronic and magnetic properties}

\begin{figure*}[htb!]
	
		\centering
		\psfig{figure=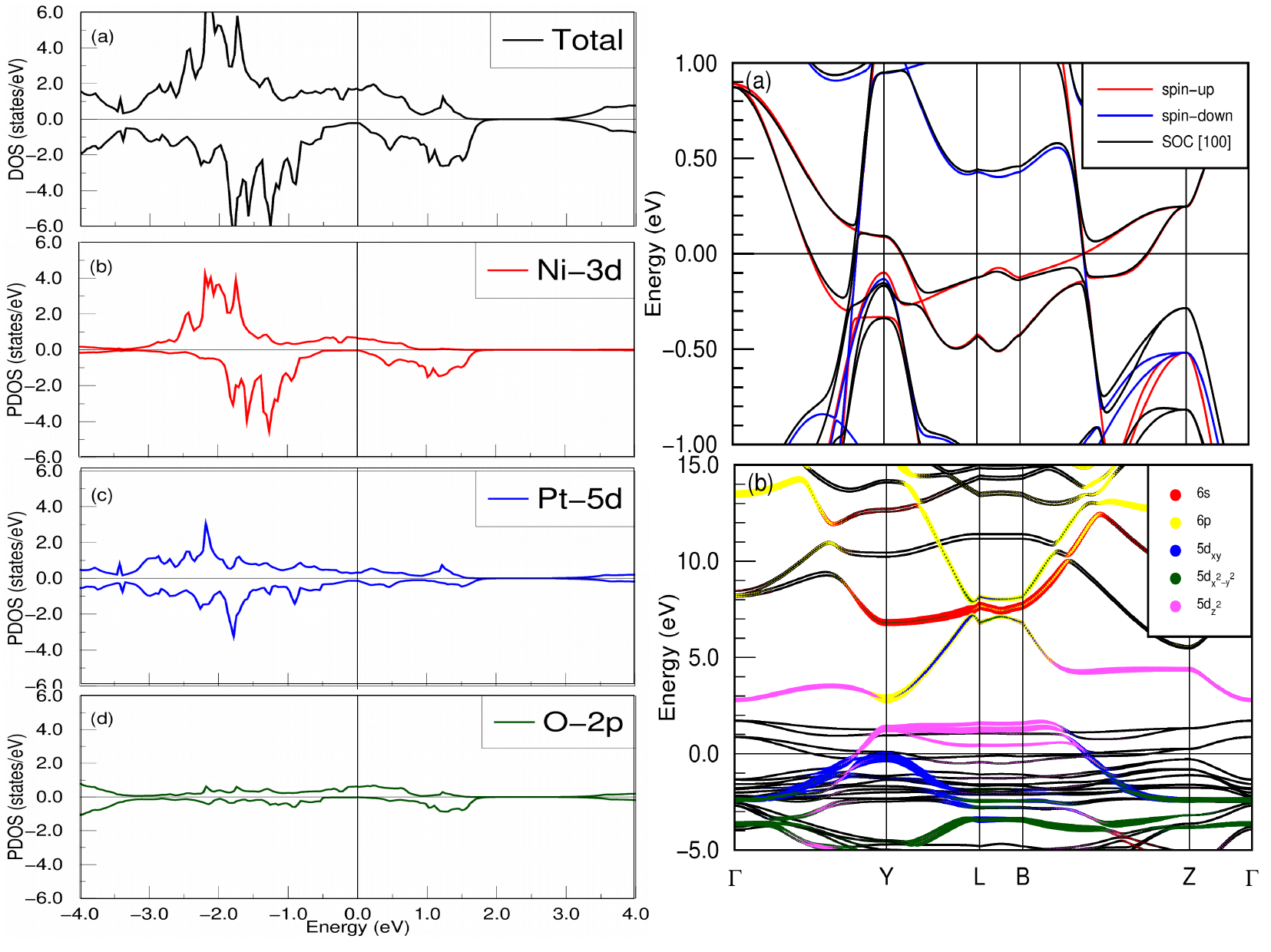,width=6.5in,height=4in} 
		\caption{Left: (a-d) Total  and partial DOS in scalar relativistic mode.  The spin-up and spin-down components are located on the upper and lower parts respectively. Right: (a) Electronic band structures of PtNiO$_2$ without and with SOC (easy anisotropy along [100] direction). Red and blue color represent the spin up and down bands respectively.  (b) Orbital characteristics analysis of Pt \textit{s}, \textit{p} and \textit{d} orbitals.}
		\label{dosband}

\end{figure*}
Similar to most of the 3R delafossites, Pt takes the charge state +1. The outermost orbital of it contains \textit{d} orbital characters, with 5\textit{d$^9$} configuration. The transition metal normally is in +3 oxidation state with 3\textit{d$^7$} configuration, lying mostly in valence band region as observed in Fig. \ref{dosband}(b). O-2\textit{p} states consisting of four electrons in the \textit{p} orbitals have dominant contribution around the $E_{F}$. Electronic structure shown in Fig. \ref{dosband}(a) supports that PtNiO$_2$ is metallic in nature due to the presence of significant density of states (DOS) around the $E_{ F}$. From our total energy calculation for non-magnetic (NM), ferro-magnetic (FM) and ferri-magnetic (FIM) configurations, ground state is found to be FM ($\uparrow$ $\uparrow$) with the lowest energy. The calculated effective magnetic moments per unit cell is 1.01 $\mu_B$.

To determine the magnetic easy axis, the full-relativistic total energies are computed along magnetization axis [001] and [100], respectively. Magnetic easy axis was found along the [100] magnetization direction with magnetic anisotropic energy of  62 meV  per unit cell. 

Electronic properties of PtNiO$_2$ resembles with its sister compound  PtCoO$_2$ reported earlier \cite{PhysRevMaterials.3.045002}. Co has six electrons on its \textit{d} orbital. Crystal field due to oxygen octahedron splits Co  \textit{d} subshell  into lower $t_{2g}$  and higher $e_{g}$ orbital, resulting in the configuration $t_{2g}^{3\uparrow}$$e_{g}^{0\uparrow}$$t_{2g}^{3\downarrow}$$e_{g}^{0\downarrow}$. With the fully occupied  $t_{2g}$ orbitals, all the magnetic moment cancels in PtCoO$_2$  whereas in PtNiO$_2$, Ni has an additional spin up electron in  $e_{g}$ orbital following the Hund's rule resulting in  $t_{2g}^{3\uparrow}$$e_{g}^{1\uparrow}$$t_{2g}^{3\downarrow}$$e_{g}^{0\downarrow}$. This renders 1 $\mu_B$ magnetic moment in an ionic picture, in agreement with our DFT calculations. 
The NiO$_6$ octahedron pushes spin-up channel away from the $E_{F}$, resulting on large exchange energy ($\approx$1 eV) [Fig. \ref{dosband}(a) right panel].  Pt forms a triangular lattice in PtNiO$_2$. Pt 5$d_{z^2}$ orbitals contribute to the conduction as depicted in [Fig. \ref{dosband}(b) right panel].  High conductivity of PtNiO$_2$ is possible due to the triangular lattice layer of Pt, where each Pt has a free electron \cite{PhysRevMaterials.3.045002,doi:10.1021/cm703404e}, as it's sister compound  PtCoO{$_2$}.

The special feature realized in a kagome-like electronic structure is a quadratic band crossing between dispersive  and non-dispersive band at $\Gamma$ \cite{du2017quadratic,sun2009topological,kang2020topological}. Quadratic band crossing points are protected by time-reversal and C$_4$ or C$_6$ rotational symmetries in different lattice models \cite{sun2009topological}. In kagome system, it is protected by time-reversal and  C$_6$ rotational symmetry \cite{du2017quadratic}. Quadratic  band dispersion are even realized in magnetic semi-metals \cite{fang2012multi}. At  K high symmetry point, Pt \textit{s+p$_x$/p$_y$} and Pt 3\textit{d$_{z^2-r^2}$} +\textit{d$_{xy}$/\textit{d$_{x^2-y^2}$}} formed a nearly three-fold degeneracy, yielding PtCoO$_2$, a kagome system hidden in the triangular lattice. The sharp band crossing the $E_{ F}$ has mainly the \textit{d} orbital character with hidden combination of \textit{s+p$_x$/p$_y$} \cite{PhysRevMaterials.3.045002}, which renders the high conductivity to PtCoO$_2$. Similar to this,  for PtNiO$_2$, orbital analysis of Pt ($s$, $p$ and $d$ orbital) reveals a quadratic band crossing of a flat band and a  dispersive band at $\Gamma$, touching at around 8 eV above the $E_{ F}$ (see Fig. \ref{dosband}(b) right panel). It shows that Pt \emph{s, p} orbitals forms a kagome-like electronic structure much above the $E_{ F}$. Such system with kagome-like structure exhibits a (near) three fold degeneracy at K-point similar to PtCoO$_2$ \cite{PhysRevMaterials.3.045002}.  This degenerate point is at around 8 eV above the $E_{ F}$, in PtNiO$_2$, representing our system as a hidden kagome-like system. This system provides a platform to understand the interplay of magnetism, hidden kagome-like lattice and Weyl fermions. 	

\subsection{Weyl properties with SOC}
\begin{figure}[h]
		\centering
		\psfig{figure=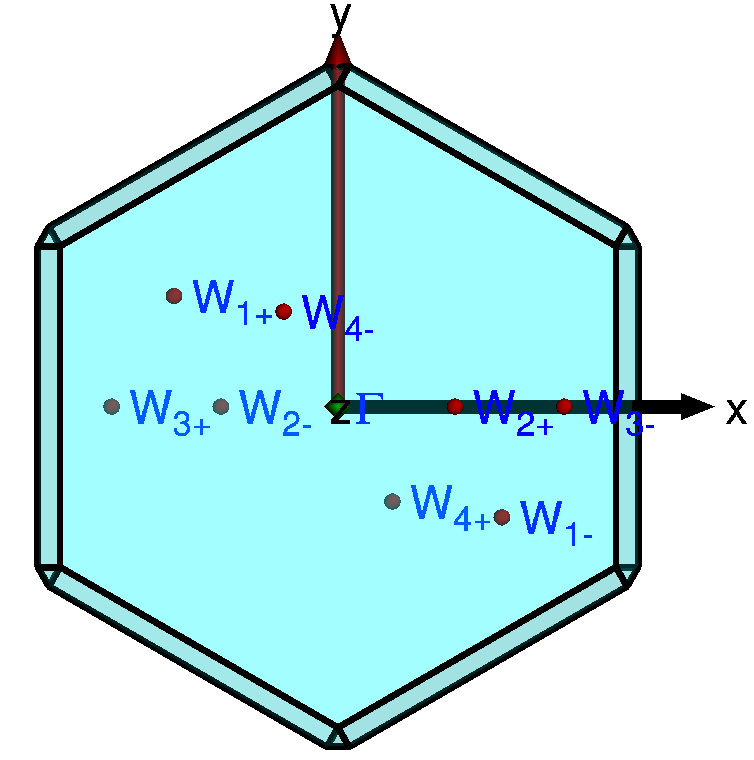,width=2in,height=2in} 
		\psfig{figure=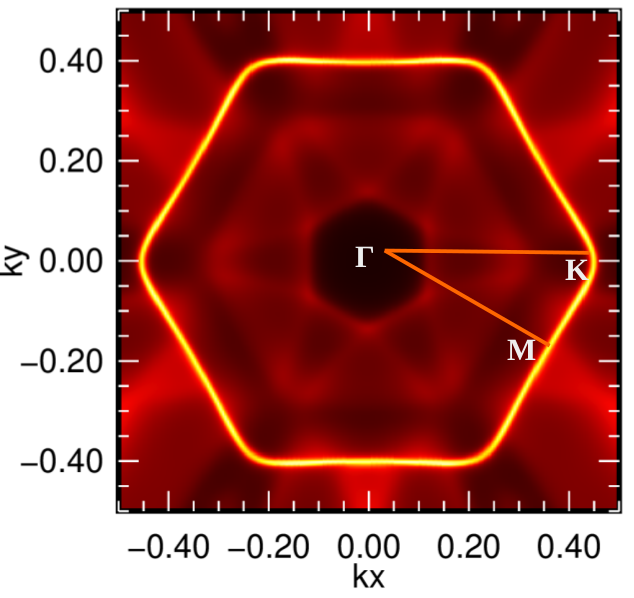,width=2in,height=2in}
		\caption{Left: Distribution of Weyl points in first 2D BZ. Right: Fermi surface spectral function. }
		\label{Weyl}
	\end{figure}

Electronic band structure of PtNiO$_2$ without and with SOC is shown in [Fig.  \ref{dosband}(a) right panel]. It can be noted that a number of bands cross each other at $E_{ F}$. Crystal field is found to split Pt in three energy states. The conduction band belonging to  2D E irreducible representation forms a crossing near high symmetry point Y slightly above $E_{ F}$. This degenerate band at $\Gamma$ splits due to reduction in symmetry along the line $\Gamma$-Y.  Along the $\Gamma$-Y and B-Z lines at the $E_{ F}$, we observe band inversion as well as band crossing. With SOC taken into account, the band structure around the $E_{ F}$ changes considerably. A significant band gap opening has been observed. The effect of SOC indicates the topological characteristics of the material. Band gap opening are noted between $\Gamma$-Y and B-Z region in the band structure. Effect of SOC results in the interactions between band topology and materials magnetism. Basing on these observation, we further explore the Weyl features.
WPs has been computed for the magnetization easy axis [100], wherein 20 pairs of WPs are found. Table \ref{table} lists their energies, multiplicity, chirality, and location of WPs  within 100 meV above the $E_{ F}$. As the location of WPs are close to $E_{F}$, their identification may be possible from transport experiment. Fig. \ref{Weyl} shows the distribution of WPs in 2D first BZ.  Pairs of W$_2$ and W$_3$'s Weyl points are connected  by M$_\textit{y}$. Due to such WPs identification this system has possibilities for large intrinsic anomalous Hall conductivity due to the large Berry curvature because large Berry curvature  comes from the monopole of Weyl points. 
In full relativistic mode (i.e. SOC),  Fermi surface spectral
functions has been investigated. Right of
Fig. \ref{Weyl} shows the  Fermi surface spectral
function for a semi infinite slab in the $k_x$ and $k_y$ plane. 

Our  Fermi surface spectral function calculation confirmed
that PtNiO$_2$ has a hidden kagome-like lattice structure.

\begin{table*}[!h]
	\begin{center}
		\caption{Characteristics of Weyl points in the PtNiO$_2$ electronic structure, when the magnetization along [100] direction. There are four sets of Weyl nodes. Within each sets, the representative $k_i$ (i = x, y, z) positions are also given, in units of  \AA$^{-1}$}.\label{table}
		\begin{tabular}{|l|l|l|l|l|l|l|l}
			\hline
			\hline
			WP& Energy(meV)& Multiplicity& $k_x$ & $k_y$& $k_z$& $\chi$\\
			\hline
			W$_1$ &24&4& -0.327&   0.221&    0.114&  -1.0\\
			W$_2$ &88 &2&  -0.234&  -0.000&  -1.029&  -1.0\\     
			W$_3$ &95 &2&  -0.452&   0.000&  -0.512&   1.0\\ 
			W$_4$ &98 &4&  -0.108&   0.189 &  1.074 &   -1.0\\

			\hline
		\end{tabular}
	\end{center}
\end{table*}

\section{Conclusions}
In conclusions, using density functional theory we investigated the electronic, magnetic and Weyl features of PtNiO$_2$. The identification of 20 pairs of confirmed Weyl points along the easy axis [100] close to Fermi level deserves the experimental work. The change in magnetization direction of PtNiO$_2$ breaks the specified symmetries, resulting change in the number of Weyl points.
Furthermore, PtNiO$_2$ is found to be a 2D hidden kagome-like lattice. Our findings suggest that PtNiO$_2$ could be a viable platform for studying the interaction between topological properties and magnetism. They also provide a good material base for future transport and surface studies in delafossites.

\begin{acknowledgement}
The author thanks M. Richter (IFW-Dresden) and Rajendra Adhikari (Kathmandu University) for fruitful discussions. This work was funded by the University Grants Commission (UGC) Nepal, through award number CRG-78/79-S\&T-03.
M.B.N. and R.G. are supported by UGC-Nepal. G.B.A. thanks Nepal Academy of Science and Technology for the PhD fellowship. M.P.G. and G.B.A. thanks Ulrike Nitzsche for the technical assistance.
\end{acknowledgement}

\bibliography{ref}

\providecommand{\latin}[1]{#1}
\providecommand*\mcitethebibliography{\thebibliography}
\csname @ifundefined\endcsname{endmcitethebibliography}
  {\let\endmcitethebibliography\endthebibliography}{}
\begin{mcitethebibliography}{49}
\providecommand*\natexlab[1]{#1}
\providecommand*\mciteSetBstSublistMode[1]{}
\providecommand*\mciteSetBstMaxWidthForm[2]{}
\providecommand*\mciteBstWouldAddEndPuncttrue
  {\def\EndOfBibitem{\unskip.}}
\providecommand*\mciteBstWouldAddEndPunctfalse
  {\let\EndOfBibitem\relax}
\providecommand*\mciteSetBstMidEndSepPunct[3]{}
\providecommand*\mciteSetBstSublistLabelBeginEnd[3]{}
\providecommand*\EndOfBibitem{}
\mciteSetBstSublistMode{f}
\mciteSetBstMaxWidthForm{subitem}{(\alph{mcitesubitemcount})}
\mciteSetBstSublistLabelBeginEnd
  {\mcitemaxwidthsubitemform\space}
  {\relax}
  {\relax}

\bibitem[Armitage \latin{et~al.}(2018)Armitage, Mele, and
  Vishwanath]{RevModPhys.90.015001}
Armitage,~N.~P.; Mele,~E.~J.; Vishwanath,~A. \emph{Rev. Mod. Phys.}
  \textbf{2018}, \emph{90}, 015001\relax
\mciteBstWouldAddEndPuncttrue
\mciteSetBstMidEndSepPunct{\mcitedefaultmidpunct}
{\mcitedefaultendpunct}{\mcitedefaultseppunct}\relax
\EndOfBibitem
\bibitem[Xia \latin{et~al.}(2019)Xia, Jin, Zhao, Chen, Zheng, Zhao, Wang, and
  Xu]{PhysRevLett.122.057205}
Xia,~B.~W.; Jin,~Y.~J.; Zhao,~J.~Z.; Chen,~Z.~J.; Zheng,~B.~B.; Zhao,~Y.~J.;
  Wang,~R.; Xu,~H. \emph{Phys. Rev. Lett.} \textbf{2019}, \emph{122},
  057205\relax
\mciteBstWouldAddEndPuncttrue
\mciteSetBstMidEndSepPunct{\mcitedefaultmidpunct}
{\mcitedefaultendpunct}{\mcitedefaultseppunct}\relax
\EndOfBibitem
\bibitem[Wang \latin{et~al.}(2012)Wang, Sun, Chen, Franchini, Xu, Weng, Dai,
  and Fang]{PhysRevB.85.195320}
Wang,~Z.; Sun,~Y.; Chen,~X.-Q.; Franchini,~C.; Xu,~G.; Weng,~H.; Dai,~X.;
  Fang,~Z. \emph{Phys. Rev. B} \textbf{2012}, \emph{85}, 195320\relax
\mciteBstWouldAddEndPuncttrue
\mciteSetBstMidEndSepPunct{\mcitedefaultmidpunct}
{\mcitedefaultendpunct}{\mcitedefaultseppunct}\relax
\EndOfBibitem
\bibitem[Liu \latin{et~al.}(2014)Liu, Zhou, Zhang, Wang, Weng, Prabhakaran, Mo,
  Shen, Fang, Dai, Hussain, and Chen]{doi:10.1126/science.1245085}
Liu,~Z.~K.; Zhou,~B.; Zhang,~Y.; Wang,~Z.~J.; Weng,~H.~M.; Prabhakaran,~D.;
  Mo,~S.-K.; Shen,~Z.~X.; Fang,~Z.; Dai,~X.; Hussain,~Z.; Chen,~Y.~L.
  \emph{Science} \textbf{2014}, \emph{343}, 864--867\relax
\mciteBstWouldAddEndPuncttrue
\mciteSetBstMidEndSepPunct{\mcitedefaultmidpunct}
{\mcitedefaultendpunct}{\mcitedefaultseppunct}\relax
\EndOfBibitem
\bibitem[Wan \latin{et~al.}(2011)Wan, Turner, Vishwanath, and
  Savrasov]{PhysRevB.83.205101}
Wan,~X.; Turner,~A.~M.; Vishwanath,~A.; Savrasov,~S.~Y. \emph{Phys. Rev. B}
  \textbf{2011}, \emph{83}, 205101\relax
\mciteBstWouldAddEndPuncttrue
\mciteSetBstMidEndSepPunct{\mcitedefaultmidpunct}
{\mcitedefaultendpunct}{\mcitedefaultseppunct}\relax
\EndOfBibitem
\bibitem[Burkov and Balents(2011)Burkov, and Balents]{PhysRevLett.107.127205}
Burkov,~A.~A.; Balents,~L. \emph{Phys. Rev. Lett.} \textbf{2011}, \emph{107},
  127205\relax
\mciteBstWouldAddEndPuncttrue
\mciteSetBstMidEndSepPunct{\mcitedefaultmidpunct}
{\mcitedefaultendpunct}{\mcitedefaultseppunct}\relax
\EndOfBibitem
\bibitem[Xu \latin{et~al.}(2011)Xu, Weng, Wang, Dai, and
  Fang]{PhysRevLett.107.186806}
Xu,~G.; Weng,~H.; Wang,~Z.; Dai,~X.; Fang,~Z. \emph{Phys. Rev. Lett.}
  \textbf{2011}, \emph{107}, 186806\relax
\mciteBstWouldAddEndPuncttrue
\mciteSetBstMidEndSepPunct{\mcitedefaultmidpunct}
{\mcitedefaultendpunct}{\mcitedefaultseppunct}\relax
\EndOfBibitem
\bibitem[Xu \latin{et~al.}(2015)Xu, Zhang, and Zhang]{PhysRevLett.115.265304}
Xu,~Y.; Zhang,~F.; Zhang,~C. \emph{Phys. Rev. Lett.} \textbf{2015}, \emph{115},
  265304\relax
\mciteBstWouldAddEndPuncttrue
\mciteSetBstMidEndSepPunct{\mcitedefaultmidpunct}
{\mcitedefaultendpunct}{\mcitedefaultseppunct}\relax
\EndOfBibitem
\bibitem[Soluyanov \latin{et~al.}(2015)Soluyanov, Gresch, Wang, Wu, Troyer,
  Dai, and Bernevig]{soluyanov2015type}
Soluyanov,~A.~A.; Gresch,~D.; Wang,~Z.; Wu,~Q.; Troyer,~M.; Dai,~X.;
  Bernevig,~B.~A. \emph{Nature} \textbf{2015}, \emph{527}, 495--498\relax
\mciteBstWouldAddEndPuncttrue
\mciteSetBstMidEndSepPunct{\mcitedefaultmidpunct}
{\mcitedefaultendpunct}{\mcitedefaultseppunct}\relax
\EndOfBibitem
\bibitem[Burkov \latin{et~al.}(2011)Burkov, Hook, and
  Balents]{PhysRevB.84.235126}
Burkov,~A.~A.; Hook,~M.~D.; Balents,~L. \emph{Phys. Rev. B} \textbf{2011},
  \emph{84}, 235126\relax
\mciteBstWouldAddEndPuncttrue
\mciteSetBstMidEndSepPunct{\mcitedefaultmidpunct}
{\mcitedefaultendpunct}{\mcitedefaultseppunct}\relax
\EndOfBibitem
\bibitem[Zhu \latin{et~al.}(2016)Zhu, Winkler, Wu, Li, and
  Soluyanov]{PhysRevX.6.031003}
Zhu,~Z.; Winkler,~G.~W.; Wu,~Q.; Li,~J.; Soluyanov,~A.~A. \emph{Phys. Rev. X}
  \textbf{2016}, \emph{6}, 031003\relax
\mciteBstWouldAddEndPuncttrue
\mciteSetBstMidEndSepPunct{\mcitedefaultmidpunct}
{\mcitedefaultendpunct}{\mcitedefaultseppunct}\relax
\EndOfBibitem
\bibitem[Winkler \latin{et~al.}(2019)Winkler, Singh, and
  Soluyanov]{Winkler_2019}
Winkler,~G.~W.; Singh,~S.; Soluyanov,~A.~A. \emph{Chin. Phys. B} \textbf{2019},
  \emph{28}, 077303\relax
\mciteBstWouldAddEndPuncttrue
\mciteSetBstMidEndSepPunct{\mcitedefaultmidpunct}
{\mcitedefaultendpunct}{\mcitedefaultseppunct}\relax
\EndOfBibitem
\bibitem[Yan(2021)]{yan2021weyl}
Yan,~B. \emph{Science China Physics, Mechanics, and Astronomy} \textbf{2021},
  \emph{64}, 217063\relax
\mciteBstWouldAddEndPuncttrue
\mciteSetBstMidEndSepPunct{\mcitedefaultmidpunct}
{\mcitedefaultendpunct}{\mcitedefaultseppunct}\relax
\EndOfBibitem
\bibitem[Shekhar \latin{et~al.}(2015)Shekhar, Nayak, Sun, Schmidt, Nicklas,
  Leermakers, Zeitler, Skourski, Wosnitza, Liu, Chen, Schnelle, Borrmann, Grin,
  Felser, and Yan]{shekhar2015extremely}
Shekhar,~C. \latin{et~al.}  \emph{Nat. Phys.} \textbf{2015}, \emph{11},
  645--649\relax
\mciteBstWouldAddEndPuncttrue
\mciteSetBstMidEndSepPunct{\mcitedefaultmidpunct}
{\mcitedefaultendpunct}{\mcitedefaultseppunct}\relax
\EndOfBibitem
\bibitem[Kumar \latin{et~al.}(2017)Kumar, Sun, Xu, Manna, Yao, S{\"u}ss,
  Leermakers, Young, F{\"o}rster, Schmidt, Borrmann, Yan, Zeitler, Shi, Felser,
  and Shekhar]{kumar2017extremely}
Kumar,~N. \latin{et~al.}  \emph{Nature Communications} \textbf{2017}, \emph{8},
  1642\relax
\mciteBstWouldAddEndPuncttrue
\mciteSetBstMidEndSepPunct{\mcitedefaultmidpunct}
{\mcitedefaultendpunct}{\mcitedefaultseppunct}\relax
\EndOfBibitem
\bibitem[Thakur \latin{et~al.}(2020)Thakur, Vir, Guin, Shekhar, Weihrich, Sun,
  Kumar, and Felser]{doi:10.1021/acs.chemmater.9b05009}
Thakur,~G.~S.; Vir,~P.; Guin,~S.~N.; Shekhar,~C.; Weihrich,~R.; Sun,~Y.;
  Kumar,~N.; Felser,~C. \emph{Chem. Mater.} \textbf{2020}, \emph{32},
  1612--1617, PMID: 32116410\relax
\mciteBstWouldAddEndPuncttrue
\mciteSetBstMidEndSepPunct{\mcitedefaultmidpunct}
{\mcitedefaultendpunct}{\mcitedefaultseppunct}\relax
\EndOfBibitem
\bibitem[Hosur and Qi(2013)Hosur, and Qi]{hosur2013recent}
Hosur,~P.; Qi,~X. \emph{C R Phys} \textbf{2013}, \emph{14}, 857--870\relax
\mciteBstWouldAddEndPuncttrue
\mciteSetBstMidEndSepPunct{\mcitedefaultmidpunct}
{\mcitedefaultendpunct}{\mcitedefaultseppunct}\relax
\EndOfBibitem
\bibitem[Morimoto \latin{et~al.}(2016)Morimoto, Zhong, Orenstein, and
  Moore]{PhysRevB.94.245121}
Morimoto,~T.; Zhong,~S.; Orenstein,~J.; Moore,~J.~E. \emph{Phys. Rev. B}
  \textbf{2016}, \emph{94}, 245121\relax
\mciteBstWouldAddEndPuncttrue
\mciteSetBstMidEndSepPunct{\mcitedefaultmidpunct}
{\mcitedefaultendpunct}{\mcitedefaultseppunct}\relax
\EndOfBibitem
\bibitem[Zyuzin and Zyuzin(2017)Zyuzin, and Zyuzin]{PhysRevB.95.085127}
Zyuzin,~A.~A.; Zyuzin,~A.~Y. \emph{Phys. Rev. B} \textbf{2017}, \emph{95},
  085127\relax
\mciteBstWouldAddEndPuncttrue
\mciteSetBstMidEndSepPunct{\mcitedefaultmidpunct}
{\mcitedefaultendpunct}{\mcitedefaultseppunct}\relax
\EndOfBibitem
\bibitem[Sirica \latin{et~al.}(2019)Sirica, Tobey, Zhao, Chen, Xu, Yang, Shen,
  Yarotski, Bowlan, Trugman, Zhu, Dai, Azad, Ni, Qiu, Taylor, and
  Prasankumar]{PhysRevLett.122.197401}
Sirica,~N. \latin{et~al.}  \emph{Phys. Rev. Lett.} \textbf{2019}, \emph{122},
  197401\relax
\mciteBstWouldAddEndPuncttrue
\mciteSetBstMidEndSepPunct{\mcitedefaultmidpunct}
{\mcitedefaultendpunct}{\mcitedefaultseppunct}\relax
\EndOfBibitem
\bibitem[Liu \latin{et~al.}(2018)Liu, Sun, Kumar, Muechler, Sun, Jiao, Yang,
  Liu, Liang, Xu, Kroder, Süß, Borrmann, Shekhar, Wang, and
  Felsar]{liu2018giant}
Liu,~E. \latin{et~al.}  \emph{Nat. Phys.} \textbf{2018}, \emph{14},
  1125--1131\relax
\mciteBstWouldAddEndPuncttrue
\mciteSetBstMidEndSepPunct{\mcitedefaultmidpunct}
{\mcitedefaultendpunct}{\mcitedefaultseppunct}\relax
\EndOfBibitem
\bibitem[Wang \latin{et~al.}(2018)Wang, Xu, Lou, Liu, Li, Huang, Shen, Weng,
  Wang, and Lei]{wang2018large}
Wang,~Q.; Xu,~Y.; Lou,~R.; Liu,~Z.; Li,~M.; Huang,~Y.; Shen,~D.; Weng,~H.;
  Wang,~S.; Lei,~H. \emph{Nature communications} \textbf{2018}, \emph{9},
  3681\relax
\mciteBstWouldAddEndPuncttrue
\mciteSetBstMidEndSepPunct{\mcitedefaultmidpunct}
{\mcitedefaultendpunct}{\mcitedefaultseppunct}\relax
\EndOfBibitem
\bibitem[Liu \latin{et~al.}(2019)Liu, Liang, Liu, Xu, Li, Chen, Pei, Shi, Mo,
  Dudin, Kim, Cacho, Li, Sun, Yang, Liu, Parkin, Felser, and
  Chen]{doi:10.1126/science.aav2873}
Liu,~D.~F. \latin{et~al.}  \emph{Science} \textbf{2019}, \emph{365},
  1282--1285\relax
\mciteBstWouldAddEndPuncttrue
\mciteSetBstMidEndSepPunct{\mcitedefaultmidpunct}
{\mcitedefaultendpunct}{\mcitedefaultseppunct}\relax
\EndOfBibitem
\bibitem[Ghimire \latin{et~al.}(2019)Ghimire, Facio, You, Ye, Checkelsky, Fang,
  Kaxiras, Richter, and van~den Brink]{PhysRevResearch.1.032044}
Ghimire,~M.~P.; Facio,~J.~I.; You,~J.-S.; Ye,~L.; Checkelsky,~J.~G.; Fang,~S.;
  Kaxiras,~E.; Richter,~M.; van~den Brink,~J. \emph{Phys. Rev. Research}
  \textbf{2019}, \emph{1}, 032044\relax
\mciteBstWouldAddEndPuncttrue
\mciteSetBstMidEndSepPunct{\mcitedefaultmidpunct}
{\mcitedefaultendpunct}{\mcitedefaultseppunct}\relax
\EndOfBibitem
\bibitem[Sakai \latin{et~al.}(2018)Sakai, Mizuta, Nugroho, Sihombing,
  Koretsune, Suzuki, Takemori, Ishii, Nishio-Hamane, Arita, Goswami, and
  Nakatsuji]{sakai2018giant}
Sakai,~A.; Mizuta,~Y.~P.; Nugroho,~A.~A.; Sihombing,~R.; Koretsune,~T.;
  Suzuki,~M.-T.; Takemori,~N.; Ishii,~R.; Nishio-Hamane,~D.; Arita,~R.;
  Goswami,~P.; Nakatsuji,~S. \emph{Nat. Phys.} \textbf{2018}, \emph{14},
  1119--1124\relax
\mciteBstWouldAddEndPuncttrue
\mciteSetBstMidEndSepPunct{\mcitedefaultmidpunct}
{\mcitedefaultendpunct}{\mcitedefaultseppunct}\relax
\EndOfBibitem
\bibitem[Guin \latin{et~al.}(2019)Guin, Manna, Noky, Watzman, Fu, Kumar,
  Schnelle, Shekhar, Sun, Gooth, and Felser]{guinnpg}
Guin,~S.~N.; Manna,~K.; Noky,~J.; Watzman,~S.~J.; Fu,~C.; Kumar,~N.;
  Schnelle,~W.; Shekhar,~C.; Sun,~Y.; Gooth,~J.; Felser,~C. \emph{NPG Asia
  Materials} \textbf{2019}, \emph{11}, 16\relax
\mciteBstWouldAddEndPuncttrue
\mciteSetBstMidEndSepPunct{\mcitedefaultmidpunct}
{\mcitedefaultendpunct}{\mcitedefaultseppunct}\relax
\EndOfBibitem
\bibitem[K{\"u}bler and Felser(2016)K{\"u}bler, and Felser]{kubler2016weyl}
K{\"u}bler,~J.; Felser,~C. \emph{EPL (Europhysics Letters)} \textbf{2016},
  \emph{114}, 47005\relax
\mciteBstWouldAddEndPuncttrue
\mciteSetBstMidEndSepPunct{\mcitedefaultmidpunct}
{\mcitedefaultendpunct}{\mcitedefaultseppunct}\relax
\EndOfBibitem
\bibitem[Kuroda \latin{et~al.}(2017)Kuroda, Tomita, Suzuki, Bareille, Nugroho,
  Goswami, Ochi, Ikhlas, Nakayama, Akebi, Noguchi, Ishii, Inami, Ono,
  Kumigashira, Varykhalov, Muro, Koretsune, Arita, Shin, Kondo, and
  Nakatsu]{kuroda2017evidence}
Kuroda,~K. \latin{et~al.}  \emph{Nat. Mater.} \textbf{2017}, \emph{16},
  1090--1095\relax
\mciteBstWouldAddEndPuncttrue
\mciteSetBstMidEndSepPunct{\mcitedefaultmidpunct}
{\mcitedefaultendpunct}{\mcitedefaultseppunct}\relax
\EndOfBibitem
\bibitem[Yang \latin{et~al.}(2017)Yang, Sun, Zhang, Shi, Parkin, and
  Yan]{Yang_2017}
Yang,~H.; Sun,~Y.; Zhang,~Y.; Shi,~W.-J.; Parkin,~S. S.~P.; Yan,~B. \emph{New
  J. Phys.} \textbf{2017}, \emph{19}, 015008\relax
\mciteBstWouldAddEndPuncttrue
\mciteSetBstMidEndSepPunct{\mcitedefaultmidpunct}
{\mcitedefaultendpunct}{\mcitedefaultseppunct}\relax
\EndOfBibitem
\bibitem[Hirschberger \latin{et~al.}(2016)Hirschberger, Kushwaha, Wang, Gibson,
  Liang, Belvin, Bernevig, Cava, and Ong]{hirschberger2016chiral}
Hirschberger,~M.; Kushwaha,~S.; Wang,~Z.; Gibson,~Q.; Liang,~S.; Belvin,~C.~A.;
  Bernevig,~B.~A.; Cava,~R.~J.; Ong,~N.~P. \emph{Nat. Mater.} \textbf{2016},
  \emph{15}, 1161--1165\relax
\mciteBstWouldAddEndPuncttrue
\mciteSetBstMidEndSepPunct{\mcitedefaultmidpunct}
{\mcitedefaultendpunct}{\mcitedefaultseppunct}\relax
\EndOfBibitem
\bibitem[Borisenko \latin{et~al.}(2019)Borisenko, Evtushinsky, Gibson, Yaresko,
  Koepernik, Kim, Ali, van~den Brink, Hoesch, Fedorov, Haubold, Kushnirenko,
  Soldatov, Schäfer, and Cava]{borisenko2019time}
Borisenko,~S.; Evtushinsky,~D.; Gibson,~Q.; Yaresko,~A.; Koepernik,~K.;
  Kim,~T.; Ali,~M.; van~den Brink,~J.; Hoesch,~M.; Fedorov,~A.; Haubold,~E.;
  Kushnirenko,~Y.; Soldatov,~I.; Schäfer,~R.; Cava,~R.~J. \emph{Nature
  communications} \textbf{2019}, \emph{10}, 3424\relax
\mciteBstWouldAddEndPuncttrue
\mciteSetBstMidEndSepPunct{\mcitedefaultmidpunct}
{\mcitedefaultendpunct}{\mcitedefaultseppunct}\relax
\EndOfBibitem
\bibitem[Shannon \latin{et~al.}(1971)Shannon, Rogers, and
  Prewitt]{doi:10.1021/ic50098a011}
Shannon,~R.~D.; Rogers,~D.~B.; Prewitt,~C.~T. \emph{Inorg. Chem.}
  \textbf{1971}, \emph{10}, 713--718\relax
\mciteBstWouldAddEndPuncttrue
\mciteSetBstMidEndSepPunct{\mcitedefaultmidpunct}
{\mcitedefaultendpunct}{\mcitedefaultseppunct}\relax
\EndOfBibitem
\bibitem[Kushwaha \latin{et~al.}(2015)Kushwaha, Sunko, Moll, Bawden, Riley,
  Nandi, Rosner, Schmidt, Arnold, Hassinger, Kim, Hoesch, Mackenzie, and
  King]{doi:10.1126/sciadv.1500692}
Kushwaha,~P.; Sunko,~V.; Moll,~P. J.~W.; Bawden,~L.; Riley,~J.~M.; Nandi,~N.;
  Rosner,~H.; Schmidt,~M.~P.; Arnold,~F.; Hassinger,~E.; Kim,~T.~K.;
  Hoesch,~M.; Mackenzie,~A.~P.; King,~P. D.~C. \emph{Sci. Adv.} \textbf{2015},
  \emph{1}, e1500692\relax
\mciteBstWouldAddEndPuncttrue
\mciteSetBstMidEndSepPunct{\mcitedefaultmidpunct}
{\mcitedefaultendpunct}{\mcitedefaultseppunct}\relax
\EndOfBibitem
\bibitem[Shannon \latin{et~al.}(1971)Shannon, Rogers, Prewitt, and
  Gillson]{doi:10.1021/ic50098a013}
Shannon,~R.~D.; Rogers,~D.~B.; Prewitt,~C.~T.; Gillson,~J.~L. \emph{Inorg.
  Chem.} \textbf{1971}, \emph{10}, 723--727\relax
\mciteBstWouldAddEndPuncttrue
\mciteSetBstMidEndSepPunct{\mcitedefaultmidpunct}
{\mcitedefaultendpunct}{\mcitedefaultseppunct}\relax
\EndOfBibitem
\bibitem[Cerqueira(2018)]{cerqueira2018structural}
Cerqueira,~T. F.~T. Structural prediction and materials design from high
  throughput to global minima optimization methods. Ph.D.\ thesis, 2018\relax
\mciteBstWouldAddEndPuncttrue
\mciteSetBstMidEndSepPunct{\mcitedefaultmidpunct}
{\mcitedefaultendpunct}{\mcitedefaultseppunct}\relax
\EndOfBibitem
\bibitem[Daou \latin{et~al.}(2017)Daou, Fr{\'e}sard, Eyert, H{\'e}bert, and
  Maignan]{daou2017unconventional}
Daou,~R.; Fr{\'e}sard,~R.; Eyert,~V.; H{\'e}bert,~S.; Maignan,~A. \emph{Science
  and Technology of Advanced Materials} \textbf{2017}, \emph{18},
  919--938\relax
\mciteBstWouldAddEndPuncttrue
\mciteSetBstMidEndSepPunct{\mcitedefaultmidpunct}
{\mcitedefaultendpunct}{\mcitedefaultseppunct}\relax
\EndOfBibitem
\bibitem[Usui \latin{et~al.}(2019)Usui, Ochi, Kitamura, Oka, Ogura, Rosner,
  Haverkort, Sunko, King, Mackenzie, and Kuroki]{PhysRevMaterials.3.045002}
Usui,~H.; Ochi,~M.; Kitamura,~S.; Oka,~T.; Ogura,~D.; Rosner,~H.;
  Haverkort,~M.~W.; Sunko,~V.; King,~P. D.~C.; Mackenzie,~A.~P.; Kuroki,~K.
  \emph{Phys. Rev. Materials} \textbf{2019}, \emph{3}, 045002\relax
\mciteBstWouldAddEndPuncttrue
\mciteSetBstMidEndSepPunct{\mcitedefaultmidpunct}
{\mcitedefaultendpunct}{\mcitedefaultseppunct}\relax
\EndOfBibitem
\bibitem[Zhang \latin{et~al.}(2019)Zhang, Ni, Zhang, Yuan, Liu, Zou, Liao, Du,
  Narayan, Zhang, Gu, Zhu, Pi, Sanvito, Han, Zou, Shi, Wan, Savrasov, and
  Ziu]{zhang2019ultrahigh}
Zhang,~C. \latin{et~al.}  \emph{Nat. Mater.} \textbf{2019}, \emph{18},
  482--488\relax
\mciteBstWouldAddEndPuncttrue
\mciteSetBstMidEndSepPunct{\mcitedefaultmidpunct}
{\mcitedefaultendpunct}{\mcitedefaultseppunct}\relax
\EndOfBibitem
\bibitem[Koepernik and Eschrig(1999)Koepernik, and Eschrig]{PhysRevB.59.1743}
Koepernik,~K.; Eschrig,~H. \emph{Phys. Rev. B} \textbf{1999}, \emph{59},
  1743--1757\relax
\mciteBstWouldAddEndPuncttrue
\mciteSetBstMidEndSepPunct{\mcitedefaultmidpunct}
{\mcitedefaultendpunct}{\mcitedefaultseppunct}\relax
\EndOfBibitem
\bibitem[Perdew \latin{et~al.}(1996)Perdew, Burke, and
  Ernzerhof]{PhysRevLett.77.3865}
Perdew,~J.~P.; Burke,~K.; Ernzerhof,~M. \emph{Phys. Rev. Lett.} \textbf{1996},
  \emph{77}, 3865--3868\relax
\mciteBstWouldAddEndPuncttrue
\mciteSetBstMidEndSepPunct{\mcitedefaultmidpunct}
{\mcitedefaultendpunct}{\mcitedefaultseppunct}\relax
\EndOfBibitem
\bibitem[Giannozzi \latin{et~al.}(2009)Giannozzi, Baroni, Bonini, Calandra,
  Car, Cavazzoni, Ceresoli, Chiarotti, Cococcioni, Dabo, Corso, Gironcoli,
  Fabris, Fratesi, Gebauer, Gerstmann, Gougoussi, and
  Kokalj]{giannozzi2009quantum}
Giannozzi,~P. \latin{et~al.}  \emph{Journal of Physics: Condensed Matter}
  \textbf{2009}, \emph{21}, 395502\relax
\mciteBstWouldAddEndPuncttrue
\mciteSetBstMidEndSepPunct{\mcitedefaultmidpunct}
{\mcitedefaultendpunct}{\mcitedefaultseppunct}\relax
\EndOfBibitem
\bibitem[Smidstrup \latin{et~al.}(2017)Smidstrup, Stradi, Wellendorff,
  Khomyakov, Vej-Hansen, Lee, Ghosh, J\'onsson, J\'onsson, and
  Stokbro]{PhysRevB.96.195309}
Smidstrup,~S.; Stradi,~D.; Wellendorff,~J.; Khomyakov,~P.~A.;
  Vej-Hansen,~U.~G.; Lee,~M.-E.; Ghosh,~T.; J\'onsson,~E.; J\'onsson,~H.;
  Stokbro,~K. \emph{Phys. Rev. B} \textbf{2017}, \emph{96}, 195309\relax
\mciteBstWouldAddEndPuncttrue
\mciteSetBstMidEndSepPunct{\mcitedefaultmidpunct}
{\mcitedefaultendpunct}{\mcitedefaultseppunct}\relax
\EndOfBibitem
\bibitem[Jain \latin{et~al.}(2013)Jain, Ong, Hautier, Chen, Richards, Dacek,
  Cholia, Gunter, Skinner, Ceder, and Persson]{jain2013commentary}
Jain,~A.; Ong,~S.~P.; Hautier,~G.; Chen,~W.; Richards,~W.~D.; Dacek,~S.;
  Cholia,~S.; Gunter,~D.; Skinner,~D.; Ceder,~G.; Persson,~K.~A. \emph{APL
  materials} \textbf{2013}, \emph{1}, 011002\relax
\mciteBstWouldAddEndPuncttrue
\mciteSetBstMidEndSepPunct{\mcitedefaultmidpunct}
{\mcitedefaultendpunct}{\mcitedefaultseppunct}\relax
\EndOfBibitem
\bibitem[Eyert \latin{et~al.}(2008)Eyert, Frésard, and
  Maignan]{doi:10.1021/cm703404e}
Eyert,~V.; Frésard,~R.; Maignan,~A. \emph{Chem. Mater.} \textbf{2008},
  \emph{20}, 2370--2373\relax
\mciteBstWouldAddEndPuncttrue
\mciteSetBstMidEndSepPunct{\mcitedefaultmidpunct}
{\mcitedefaultendpunct}{\mcitedefaultseppunct}\relax
\EndOfBibitem
\bibitem[Du \latin{et~al.}(2017)Du, Zhou, and Fiete]{du2017quadratic}
Du,~L.; Zhou,~X.; Fiete,~G.~A. \emph{Phys. Rev. B} \textbf{2017}, \emph{95},
  035136\relax
\mciteBstWouldAddEndPuncttrue
\mciteSetBstMidEndSepPunct{\mcitedefaultmidpunct}
{\mcitedefaultendpunct}{\mcitedefaultseppunct}\relax
\EndOfBibitem
\bibitem[Sun \latin{et~al.}(2009)Sun, Yao, Fradkin, and
  Kivelson]{sun2009topological}
Sun,~K.; Yao,~H.; Fradkin,~E.; Kivelson,~S.~A. \emph{Phys. Rev. Lett.}
  \textbf{2009}, \emph{103}, 046811\relax
\mciteBstWouldAddEndPuncttrue
\mciteSetBstMidEndSepPunct{\mcitedefaultmidpunct}
{\mcitedefaultendpunct}{\mcitedefaultseppunct}\relax
\EndOfBibitem
\bibitem[Kang \latin{et~al.}(2020)Kang, Fang, Ye, Po, Denlinger, Jozwiak,
  Bostwick, Rotenberg, Kaxiras, Checkelsky, and Comin]{kang2020topological}
Kang,~M.; Fang,~S.; Ye,~L.; Po,~H.~C.; Denlinger,~J.; Jozwiak,~C.;
  Bostwick,~A.; Rotenberg,~E.; Kaxiras,~E.; Checkelsky,~J.~G.; Comin,~R.
  \emph{Nat. Commun.} \textbf{2020}, \emph{11}, 4004\relax
\mciteBstWouldAddEndPuncttrue
\mciteSetBstMidEndSepPunct{\mcitedefaultmidpunct}
{\mcitedefaultendpunct}{\mcitedefaultseppunct}\relax
\EndOfBibitem
\bibitem[Fang \latin{et~al.}(2012)Fang, Gilbert, Dai, and
  Bernevig]{fang2012multi}
Fang,~C.; Gilbert,~M.~J.; Dai,~X.; Bernevig,~B.~A. \emph{Phys. Rev. Lett.}
  \textbf{2012}, \emph{108}, 266802\relax
\mciteBstWouldAddEndPuncttrue
\mciteSetBstMidEndSepPunct{\mcitedefaultmidpunct}
{\mcitedefaultendpunct}{\mcitedefaultseppunct}\relax
\EndOfBibitem
\end{mcitethebibliography}

\end{document}